\documentclass[12pt]{iopart}
\usepackage{graphicx,amssymb,hyperref}
\usepackage{cmap}
\usepackage[utf8]{inputenc}
\usepackage[T1]{fontenc}

\def\imo{i}

\def\K{{\cal K}}
\let\s\varsigma
\def\Order#1{{\cal O}\left(#1\right)}
\DeclareMathAlphabet{\pazocal}{OMS}{zplm}{m}{n}
\def\dfrac#1#2{\frac{\displaystyle #1}{\displaystyle #2}}

\begin{document}
\title[Analytic expressions for quasinormal modes and grey-body factors]{Analytic expressions for quasinormal modes and grey-body factors in the eikonal limit and beyond}
\author{R A Konoplya$^\dag$ and A Zhidenko$^\ddag$}
\address{$^\dag$ Research Centre for Theoretical Physics and Astrophysics, \\ Institute of Physics, Silesian University in Opava, \\ Bezručovo náměstí 13, CZ-74601 Opava, Czech Republic}
\ead{roman.konoplya@gmail.com}
\address{$^\ddag$ Centro de Matemática, Computação e Cognição (CMCC),\\ Universidade Federal do ABC (UFABC), \\ Rua Abolição, CEP: 09210-180, Santo André, SP, Brazil}
\ead{olexandr.zhydenko@ufabc.edu.br}

\begin{abstract}
Although the WKB series converges only asymptotically and guarantees the exact result solely in the eikonal regime, we have managed to derive concise analytical expressions for the quasinormal modes and grey-body factors of black holes, expanding beyond the eikonal approximation. Remarkably, these expressions demonstrate unexpectedly strong accuracy. We suggest a comprehensive approach for deriving analytical expressions for grey-body factors and quasinormal modes at various orders beyond the eikonal approximation. Two cases are examined as examples: the Schwarzschild-de Sitter black hole and hairy black holes within the framework of Effective Field Theory. We have publicly shared a generic code that calculates analytical expressions for grey-body factors and quasinormal modes of spherical black holes.
\end{abstract}

\maketitle

\section{Introduction}

The proper oscillation frequencies of four and higher-dimensional black holes, referred to as {\it quasinormal modes} \cite{Konoplya:2011qq, Cardoso:2008bp, Kokkotas:1999bd}, generally follow intricate differential equations that are challenging to solve exactly. Consequently, numerical data for the frequencies is commonly obtained. However, in specific scenarios, various approximations permit analytical treatment. One such instance is presented by near-extreme asymptotically de Sitter black holes \cite{Cardoso:2003sw, Churilova:2021nnc}, as well as Einstein-Gauss-Bonnet and Einstein-Lovelock black holes at specific fixed values of the coupling constant (see, for example, \cite{Gonzalez:2017gwa} and references therein).

Arguably the most renowned case of analytically addressing the spectral problem for black holes involves the high real oscillation frequency limit, corresponding to high values of the multipole number $\ell$ \cite{Mashhoon:1982im}. In this scenario, the well-known formula was derived using either the inverted Pöschl-Teller potential \cite{Mashhoon:1982im} or the first-order WKB approach \cite{Schutz:1985km}:
\begin{equation}
\omega = \frac{1}{{3 \sqrt{3} M}} \left(\ell+\frac{1}{2} - \imo \left(n+\frac{1}{2}\right)\right) + \Order{\frac{1}{\ell}},
\end{equation}
Here, $M$ represents the mass of the black hole, and $n$ stands for the overtone number.

There are several reasons why this regime has garnered significant attention. First and foremost, the aforementioned eikonal formula offers a robust approximation not only for $\ell \gg 1$, but also for moderate values of the multipole number. In this regime, the spectrum becomes independent of the spin field. Additionally, it has been observed that a correlation exists between the eikonal quasinormal modes and the characteristics of null geodesics \cite{Cardoso:2008bp, Konoplya:2017wot, Konoplya:2022gjp}, and consequently, with the shadows cast by black holes \cite{Jusufi:2020dhz,Cuadros-Melgar:2020kqn}. Moreover, there are indications that both the eikonal regime and null geodesics may possess a holographic interpretation \cite{Berenstein:2020vlp, Hadar:2022xag, Konoplya:2020ptx}. After all, the eikonal regime can bring qualitatively new features to the black hole spectrum, leading even to the instability at high $\ell$, called \textit{the eikonal instability} \cite{Takahashi:2010gz,Konoplya:2017lhs}.

Recent extensive research has been carried out on the eikonal limit of black hole quasinormal spectra in various theories of gravity \cite{Churilova:2019jqx, Glampedakis:2019dqh, Silva:2019scu,Bolokhov:2022rqv,Malybayev:2021lfq}. However, to the best of our knowledge, there has been no comparable study focusing on grey-body factors. These factors enable the investigation of the portion of Hawking radiation that is reflected by the potential barrier back to the event horizon. The same WKB approach used to determine quasinormal modes \cite{Schutz:1985km, Iyer:1986np, Konoplya:2003ii, Konoplya:2019hlu} can be effectively applied to derive grey-body factors \cite{Konoplya:2023bpf,Konoplya:2020jgt,Li:2021epb,Gogoi:2023fow,Konoplya:2023ppx}.

In the present study, we develop comprehensive tools to derive analytical expressions for the quasinormal modes of black holes beyond the eikonal approximation. Notably, to maintain a consistent level of accuracy, achieving higher orders of precision in the $\ell^{-1}$ expansion necessitates corresponding increases in the orders of the WKB approach itself. As an illustration, we present a concise analytical expression for the quasinormal modes of fields with various spins for Schwarzschild-de Sitter black holes at the order of $\ell^{-6}$. Furthermore, we deduce the analytic eikonal formula for the grey-body factors of Schwarzschild-de Sitter black holes and extend it to higher orders beyond the eikonal regime. Demonstrating its applicability even at low multipole numbers $\ell$, we establish that the derived analytical expressions offer reasonable approximations.

Additionally, we provide an automatic Mathematica\textregistered{} code that facilitates the derivation of analytical formulas for quasinormal modes and grey-body factors in the eikonal order, and at orders beyond the eikonal.

The structure of the paper unfolds as follows. Section~\ref{sec:WKB} delves into the utilization of the WKB technique to ascertain quasinormal modes beyond the eikonal regime. In Section~\ref{sec:examples}, we apply this technique to obtain analytical expressions for the quasinormal modes of both Schwarzschild-de Sitter black holes and black holes within the Effective Field Theory framework. Section~\ref{sec:greybody} formulates a comprehensive approach to derive analytical expressions for grey-body factors within the eikonal limit, and including higher-order corrections represented by an expansion in powers of $\ell^{-1}$. Finally, we conclude by summarizing the obtained results and highlighting some unresolved issues.

\section{WKB formula}\label{sec:WKB}

When the effective potential $V(r_*)$ in the wavelike equation,
\begin{equation}\label{wavelike}
\frac{d^2\Psi}{dr_*^2}+(\omega^2-V(r_*))\Psi(r_*)=0,
\end{equation}
has a form of a barrier with a single peak, the WKB formula is appropriate for obtaining the dominant quasinormal modes, satisfying the boundary conditions,
\begin{equation}\label{boundaryconditions}
\Psi(r_*\to\pm\infty)\propto e^{\pm\imo \omega r_*},
\end{equation}
which are purely ingoing wave at the horizon ($r_*\to-\infty$) and purely outgoing wave at spatial infinity or cosmological horizon ($r_*\to\infty$).

The WKB method is based on matching of the asymptotic solutions, which satisfy the quasinormal boundary conditions (\ref{boundaryconditions}), with the Taylor expansion around the peak of the potential barrier. The first-order WKB formula represents the eikonal approximation and is exact in the limit $\ell \rightarrow \infty$. Then, the general WKB expression for the frequencies can be written in the form of expansion around the eikonal limit as follows \cite{Konoplya:2019hlu}:
\begin{eqnarray}\label{WKBformula-spherical}
\omega^2&=&V_0+A_2(\K^2)+A_4(\K^2)+A_6(\K^2)+\ldots\\\nonumber&-&\imo \K\sqrt{-2V_2}\left(1+A_3(\K^2)+A_5(\K^2)+A_7(\K^2)\ldots\right),
\end{eqnarray}
and the matching conditions for the quasinormal modes imply that
\begin{equation}
\K=n+\frac{1}{2}, \quad n=0,1,2,\ldots,
\end{equation}
where $n$ is the overtone number, $V_0$ is the value of the effective potential in its maximum, $V_2$ is the value of the second derivative of the potential in this point, and $A_i$ for $i=2, 3, 4, \ldots$ is $i-th$ WKB order correction term beyond the eikonal approximation which depends on $\K$ and derivatives of the potential in its maximum up to the order $2i$. The explicit form of $A_i$ can be found in \cite{Iyer:1986np} for the second and third WKB order, in \cite{Konoplya:2003ii} for the 4th-6th orders and in \cite{Matyjasek:2017psv} for the 7th-13th orders.

For the test fields and gravitational perturbations in a spherically symmetric background, the effective potential can be represented as follows:
\begin{equation}\label{potential-multipole}
V(r_*)=\kappa^2\left(H(r_*)+\Order{\kappa^{-1}}\right),
\end{equation}
where $\kappa\equiv\ell+\frac{1}{2}$ and $\ell=s,s+1,s+2,\ldots$ is the positive half(integer) multipole number, which has minimal value equal to the spin of the perturbations $s$. Notice that from here on, when discussing expansion beyond the eikonal limit, we will use an expansion in terms of powers of $\kappa^{-1}$ instead of $\ell^{-1}$ because the resulting formulas appear more concise this way.

The function $H(r_*)$ usually has the form of the single peak, therefore, the point of maximum of the potential (\ref{potential-multipole}) can be expanded in series
\begin{equation}\label{rmax}
r_{max}=r_0+r_1\kappa^{-1}+r_2\kappa^{-2}+\ldots.
\end{equation}

Substituting (\ref{rmax}) into the eikonal formula
\begin{eqnarray}
\omega&=&\sqrt{V_0-\imo \K\sqrt{-2V_2}},
\end{eqnarray}
and expanding the result in terms of $\kappa^{-1}$, one can derive the eikonal formulas,
\begin{eqnarray}\label{eikonal-formulas}
\omega=\Omega\kappa-\imo\lambda\K+\Order{\kappa^{-1}},
\end{eqnarray}
which provides a good approximation for $\kappa\gg\K$.

In \cite{Cardoso:2008bp}, a proposition was put forth asserting that parameters related to unstable circular null geodesics encircling a stationary, spherically symmetric, and asymptotically flat or de Sitter black hole are dual to the quasinormal modes emitted by the black hole in the $\ell \gg n$ regime. Here, $\ell$ denotes the multipole number, and $n$ is the overtone. Specifically, the real and imaginary components of the $\ell \gg n$ quasinormal mode are proportional to the frequency and instability timescale of the circular null geodesics, respectively, as given by:
\begin{equation}\label{QNM}
\omega_n=\Omega\ell-\imo(n+1/2)|\lambda|, \quad \ell \gg n.
\end{equation}
In this context, $\Omega$ stands for the angular velocity at the unstable null geodesics, and $\lambda$ represents the Lyapunov exponent.

However, while this correspondence holds true for the Schwarzschild black hole and certain other cases, \cite{Konoplya:2017wot} demonstrated that the correspondence breaks down when the wave-like equation lacks a dominant eikonal limit governed by the centrifugal term $f(r) \ell (\ell +1)/r^2$. This scenario arises, for instance, in gravitational perturbations within the Einstein-Gauss-Bonnet \cite{Konoplya:2017wot,Konoplya:2020bxa} or Einstein-dilaton-Gauss-Bonnet theories \cite{Konoplya:2019hml}. More broadly, the mentioned correspondence operates when the first-order WKB formula \cite{Schutz:1985km} holds, implying an effective potential with a solitary maximum and two turning points.

Nonetheless, even within such cases, the WKB formula might fail to reproduce the entire spectrum within the eikonal regime, as witnessed in the Schwarzschild-de Sitter spacetime. In this instance, two branches of modes emerge: Schwarzschild modes altered by the $\Lambda$-term \cite{Zhidenko:2003wq,Konoplya:2004uk}, and de Sitter modes modified by the presence of a black hole \cite{Konoplya:2022xid,Cardoso:2017soq}. The latter branch comprises purely imaginary (non-oscillating) modes, which evade representation through the WKB formula \cite{Konoplya:2022gjp}. Consequently, while the correspondence indicated by equation (\ref{QNM}) formally stands, the parameter $n$ no longer corresponds to the correct overtone number. This discrepancy arises because only a segment of the spectrum $\ell \gg n$ is encompassed by the correspondence.

We notice, that expansion (\ref{rmax}) can be performed up to arbitrary order, then we can derive the higher-order corrections to the eikonal formula (\ref{eikonal-formulas}) by using the higher-order WKB formula. If we employ the WKB formula of the order $n$ beyond the eikonal approximation we can determine the correction of the order $\kappa^{1-n}$ to the analytic expression (\ref{eikonal-formulas}). In particular, the 7th-order WKB formula \cite{Matyjasek:2017psv} allows us to find the expansion up to the order of $\kappa^{-6}$.

\section{Approximate analytic formulas for quasinormal modes}\label{sec:examples}
\subsection{Schwarzschild-de Sitter black hole}
As a warm-up example, we will consider the Schwarzschild-de Sitter black hole, which is given by the line element
\begin{equation}\label{SdS-metric}
ds^2 = -f(r)dt^2 + \frac{dr^2}{f(r)} + r^2 (d\theta^2 + \sin^2\theta\phi^2),
\end{equation}
\begin{equation}
f(r) = 1 - \frac{2M}{r} - \frac{\Lambda r^2}{3}.
\end{equation}

The general covariant equations for the scalar ($\Phi$), electromagnetic ($A_\mu$), and Dirac ($\Upsilon$) and gravitational fields can be written in the following way:
\begin{equation}\label{perturb}
\begin{array}{lcl}
\dfrac{1}{\sqrt{-g}}\partial_\mu \left(\sqrt{-g}g^{\mu \nu}\partial_\nu\Phi\right)&=&0,
\\
\dfrac{1}{\sqrt{-g}}\partial_{\mu} \left(F_{\rho\sigma}g^{\rho \nu}g^{\sigma \mu}\sqrt{-g}\right)&=&0\,,
\\
\gamma^{\alpha} \left( \dfrac{\partial}{\partial x^{\alpha}} - \Gamma_{\alpha} \right) \Upsilon&=&0,
\\
\delta R_{\mu \nu}-\Lambda \delta g_{\mu \nu}&=&0,
\end{array}
\end{equation}
where $F_{\mu\nu}=\partial_\mu A_\nu-\partial_\nu A_\mu$ is the electromagnetic tensor, $\gamma^{\alpha}$ are gamma matrices, $\Gamma_{\alpha}$ are spin connections in the tetrad formalism, and $R_{\mu \nu}$ is the Ricci tensor.

After separation of the variables the above equations (\ref{perturb}) take the Schrödinger wave-like form (\ref{wavelike}) with the effective potential, which has the following form for the bosonic cases (see, for instance, a review \cite{Konoplya:2011qq}):
\begin{equation}\label{SdS-potantial}
V(r) = f(r)\left(\frac{\kappa^2-1/4}{r^2} + \frac{1-s^2}{r}f'(r)-s(s-1)\Lambda\right),
\end{equation}
where $s=0$ describes the minimally coupled test scalar field, $s=1$ corresponds to the Maxwell field, and $s=2$ provides the equation for the axial-type (odd) gravitational perturbations, and the tortoise coordinate is defined as
\begin{equation}\label{SdS-tortoise}
dr_*=\frac{dr}{f(r)}.
\end{equation}

The even gravitational perturbations are iso-spectral with the odd ones, so here we will consider only the simpler, odd type of perturbations. The maximum of the effective potential is located at
\begin{eqnarray}\label{SdSmax}
r_{max}=3M+M\frac{(s-1)\s^2(3 + (s-2) \s^2)}{\kappa^2}+\Order{\frac{1}{\kappa^4}},
\end{eqnarray}
where \mbox{$\s\equiv\sqrt{1-9\Lambda M^2}$},
and, using the third-order WKB formula, we obtain a generalization of the formula (19) in \cite{Zhidenko:2003wq}:
\begin{eqnarray}\label{SdSspectrum}
\omega=\frac{\s}{3\sqrt{3}M}\left(\kappa+\frac{\s^2 - 108 + 72s (3 - 3s + (s-3)\s^2)- 60\s^2\K^2}{432 \kappa}\right)-
\\\nonumber\hspace{-70pt}
\frac{\imo\K\s}{3\sqrt{3}M}\left(1+\s^2 \frac{864 s \left(3(s-5)-5(s-3)\s^2\right) + 8568 - 8255\s^2 + 940\s^2\K^2}{15552\kappa^2}\right) + \Order{\frac{1}{\kappa^3}}.
\end{eqnarray}

For the Dirac field ($s=1/2$) one has two iso-spectral potentials
\begin{equation}
V_{\pm}(r) = W^2\pm\frac{dW}{dr_*}, \quad W\equiv \kappa\frac{\sqrt{f(r)}}{r}.
\end{equation}
The iso-spectral wave functions can be transformed one into another by the Darboux transformation
\begin{equation}\label{psi}
\Psi_{+}=q \left(W+\frac{d}{dr_*}\right) \Psi_{-}, \quad q=const,
\end{equation}
so that it is sufficient to consider quasinormal modes and grey-body factors for only one of the effective potentials.

The maxima of the effective potentials have the following expansions:
\begin{equation}\label{SdSmaxDirac}
r_{max}=3M \mp \frac{\sqrt{3}M\s}{2\kappa}\left(1-\frac{3}{8\kappa^2}-\frac{\s^2}{12 \kappa^2}\right)+\Order{\frac{1}{\kappa^4}},
\end{equation}
so that all the odd powers of $\kappa$ have opposite signs. After substituting (\ref{SdSmaxDirac}) into the third-order WKB formula the opposite signs cancel out, and we obtain the following formula for any of the Dirac potentials:

\begin{eqnarray}\label{SdSspectrumDirac}
\omega&=&\frac{\s}{3\sqrt{3}M}\left(\kappa - \frac{7\s^2}{432\kappa} - \frac{5\K^2\s^2}{36\kappa}\right)
\\\nonumber&&
-\frac{\imo\K\s}{3\sqrt{3}M}\left(1 - \frac{7\s^2}{216\kappa^2} + \frac{385\s^4}{15552\kappa^2} + \frac{235\K^2\s^4}{3888\kappa^2}\right) + \Order{\frac{1}{\kappa^3}}.
\end{eqnarray}

\begin{table}
\begin{flushright}
\begin{tabular}{|l|c|c|c|}
\hline
 $i$& $\ell=0$ ($\kappa=1/2$)  &  $\ell=1$ ($\kappa=3/2$) & $\ell=2$ ($\kappa=5/2$)  \\
\hline
  $0$&$0.096225-0.096225 \imo$ & $0.288675-0.096225 \imo$ & $0.481125-0.096225 \imo$ \\
  $1$&$0.108699-0.096225 \imo$ & $0.292833-0.096225 \imo$ & $0.483620-0.096225 \imo$ \\
  $2$&$0.108699-0.109788 \imo$ & $0.292833-0.097732 \imo$ & $0.483620-0.096768 \imo$ \\
  $3$&$0.111895-0.109788 \imo$ & $0.292951-0.097732 \imo$ & $0.483646-0.096768 \imo$ \\
  $4$&$0.111895-0.104772 \imo$ & $0.292951-0.097670 \imo$ & $0.483646-0.096760 \imo$ \\
  $5$&$0.105203-0.104772 \imo$ & $0.292924-0.097670 \imo$ & $0.483643-0.096760 \imo$ \\
  $6$&$0.105203-0.088640 \imo$ & $0.292924-0.097648 \imo$ & $0.483643-0.096758 \imo$ \\
\hline
exact&$0.110455-0.104896 \imo$ & $0.292936-0.097660 \imo$ & $0.483644-0.096759 \imo$ \\
\hline
\end{tabular}
\end{flushright}
\caption{Dominant ($\K=1/2$) quasinormal modes of the scalar field ($s=0$) in the Schwarzschild background ($M=1$), calculated the approximate formula of the order $i$ in terms of $\kappa^{-1}$, and the accurate values \cite{Zhidenko:2006rs} for comparison.}\label{tabl:QNMSscalar}
\end{table}

\begin{table}
\begin{flushright}
\begin{tabular}{|l|c|c|c|}
\hline
 $i$& $\ell=1$ ($\kappa=3/2$)  &  $\ell=2$ ($\kappa=5/2$) &  $\ell=3$ ($\kappa=7/2$)\\
\hline
  $0$&$0.288675-0.096225 \imo$ & $0.481125-0.096225 \imo$ & $0.673575-0.096225 \imo$\\
  $1$&$0.250066-0.096225 \imo$ & $0.457960-0.096225 \imo$ & $0.657029-0.096225 \imo$\\
  $2$&$0.250066-0.092980 \imo$ & $0.457960-0.095057 \imo$ & $0.657029-0.095629 \imo$\\
  $3$&$0.248454-0.092980 \imo$ & $0.457612-0.095057 \imo$ & $0.656902-0.095629 \imo$\\
  $4$&$0.248454-0.092627 \imo$ & $0.457612-0.095011 \imo$ & $0.656902-0.095617 \imo$\\
  $5$&$0.248232-0.092627 \imo$ & $0.457594-0.095011 \imo$ & $0.656899-0.095617 \imo$\\
  $6$&$0.248232-0.092479 \imo$ & $0.457594-0.095004 \imo$ & $0.656899-0.095616 \imo$\\
\hline
exact&$0.248263-0.092488 \imo$ & $0.457596-0.095004 \imo$ & $0.656899-0.095616 \imo$\\
\hline
\end{tabular}
\end{flushright}
\caption{Dominant ($\K=1/2$) quasinormal modes of the electromagnetic perturbations ($s=1$) of the Schwarzschild black hole ($M=1$), calculated the approximate formula of the order $i$ in terms of $\kappa^{-1}$, and the accurate values \cite{Konoplya:2004uk} for comparison.}\label{tabl:QNMSMaxwell}
\end{table}

\begin{table}
\begin{flushright}
\begin{tabular}{|l|c|c|}
\hline
  $i$& $\ell=2$ ($\kappa=5/2$) & $\ell=3$ ($\kappa=7/2$)  \\
\hline
  $0$&$0.481125-0.096225 \imo$ & $0.673575-0.096225 \imo$ \\
  $1$&$0.380980-0.096225 \imo$ & $0.602043-0.096225 \imo$ \\
  $2$&$0.380980-0.089925 \imo$ & $0.602043-0.093011 \imo$ \\
  $3$&$0.374036-0.089925 \imo$ & $0.599512-0.093011 \imo$ \\
  $4$&$0.374036-0.088671 \imo$ & $0.599512-0.092684 \imo$ \\
  $5$&$0.373642-0.088671 \imo$ & $0.599439-0.092684 \imo$ \\
  $6$&$0.373642-0.088716 \imo$ & $0.599439-0.092690 \imo$ \\
\hline
exact&$0.373671-0.088962 \imo$ & $0.599444-0.092703 \imo$ \\
\hline
\end{tabular}
\end{flushright}
\caption{Dominant ($\K=1/2$) quasinormal modes of the gravitational perturbations ($s=2$) of the Schwarzschild black hole ($M=1$), calculated the approximate formula of the order $i$ in terms of $\kappa^{-1}$, and the accurate values \cite{Leaver:1985ax} for comparison.}\label{tabl:QNMSgrav}
\end{table}

\begin{table}
\begin{flushright}
\begin{tabular}{|l|c|c|}
\hline
  $i$&$\ell=1/2$ ($\kappa=1$) & $\ell=3/2$ ($\kappa=2$) \\
\hline
  $0$&$0.192450-0.096225 \imo$&$0.384900-0.096225 \imo$ \\
  $1$&$0.182649-0.096225 \imo$&$0.380000-0.096225 \imo$ \\
  $2$&$0.182649-0.096943 \imo$&$0.380000-0.096404 \imo$ \\
  $3$&$0.182925-0.096943 \imo$&$0.380034-0.096404 \imo$ \\
  $4$&$0.182925-0.096928 \imo$&$0.380034-0.096404 \imo$ \\
  $5$&$0.183038-0.096928 \imo$&$0.380038-0.096404 \imo$ \\
  $6$&$0.183038-0.097075 \imo$&$0.380038-0.096406 \imo$ \\
\hline
exact&$0.182963-0.096982 \imo$&$0.380037-0.096405 \imo$ \\
\hline
\end{tabular}
\end{flushright}
\caption{Dominant ($\K=1/2$) quasinormal modes of the massless Dirac field in the Schwarzschild background ($M=1$), calculated the approximate formula of the order $i$ in terms of $\kappa^{-1}$, and the accurate values\cite{Jing:2005dt} for comparison.}\label{tabl:QNMSDirac}
\end{table}

From the tables~\ref{tabl:QNMSscalar}-\ref{tabl:QNMSDirac} (see $i=2$) we conclude that eqs.~(\ref{SdSspectrum}) and (\ref{SdSspectrumDirac}) provide quite a good approximation for the dominant modes even for the lowest multipole number $\ell=s$, where the error is several percents. The higher orders of the expansion in terms of $\kappa^{-1}$ provide better accuracy for the dominant modes as well as for overtones\footnote{The accurate values of the quasinormal modes are available from
\url{http://qnms.way.to}.}. However, the expansion converges only asymptotically and does not improve the accuracy after some order. For the practical purpose we limit our consideration at the expansion order $i=6$, which can be obtained from the seventh-order WKB formula beyond eikonal.

\subsection{Black holes in the Effective Field Theory}

Here we will use the resultant wave like equation for the axial gravitational perturbations in the Effective Field Theory deduced in \cite{Mukohyama:2023xyf}. The deduction of the wave equation for the even gravitational as well as scalar, electromagnetic and Dirac perturbations are highly non-trivial in the Effective Field Theory, because all of them are characterized by their specific effective propagation speed and tortoise coordinate, so that we used here the case for which the wave-like equation is already known.

The function~$F\equiv dr/dr_*$ relating the tortoise and the Schwarzschild radial coordinates is
\begin{equation}\label{eq:drdrs-Hayward}
F(r)=\frac{r^4 - \mu (r^3 + \sigma^3)}{r (r^3 + \sigma^3)},
\end{equation}
where $\mu$ is a double asymptotic mass. Notice, in the Effective Field Theory this function is not the same as one could expect from the form of the metric, which is the Hayward metric \cite{Hayward:2005gi},
\begin{equation}
g_{tt} = g_{rr}^{-1}= 1- \frac{\mu r^2}{r^3 +\sigma^3}.
\end{equation}

The position of the odd-mode horizon~$r_g>0$ is given by $F(r_g)=0$, or equivalently,
\begin{equation}
\mu=\frac{r_g^4}{r_g^3+\sigma^3},
\label{eq:r_g-Hayward}
\end{equation}
which has a single positive solution so long as the mass parameter $\mu$ and the coupling $\sigma$ are positive.
The effective potential is
\begin{eqnarray}
\label{eq:Hayward_potential}
V(r) &=&\left[1-\frac{\mu(r^3+\sigma^3)}{r^4}\right] \Biggl\{\frac{(\kappa^2\!-\!1/4)r^4}{(r^3+\sigma^3)^2}
\\\nonumber&&
-\frac{3\left[4\mu r^9+2\sigma^3r^6(8r-\mu)+\sigma^6r^3(r-7\mu)-\mu\sigma^9\right]}{4(r^3+\sigma^3)^4}\Biggr\}.
\end{eqnarray}
The eikonal formula
\begin{equation}\label{eikonal-Hayward}
\omega = \frac{2\kappa}{3 \sqrt{3}r_g} \left(1+\frac{11 \sigma^3}{27}\right)-\frac{2\imo\K}{3\sqrt{3}r_g}\left(1+\frac{35 \sigma ^3}{27}\right)+\Order{\kappa^{-1},\sigma^6},
\end{equation}
was deduced in \cite{Konoplya:2023ppx}.

Using the higher-order expansion we find the generalization of the above analytic formula,
\begin{eqnarray}\label{Hayward-beyondeikonal}
\omega &=& \frac{2\kappa}{3 \sqrt{3}r_g} \left(1+\frac{11 \sigma^3}{27}-\frac{256 \sigma^6}{243}\right)-\frac{2\imo\K}{3\sqrt{3}r_g}\left(1+\frac{35 \sigma ^3}{27}-\frac{352 \sigma ^6}{243}\right)
\\\nonumber&&
-\frac{60\K^2+547}{648 \sqrt{3}\kappa r_g}-\frac{\left(5460\K^2+10321\right)\sigma^3}{17496 \sqrt{3}\kappa r_g}-\frac{32 \left(816\K^2-1073\right)\sigma^6}{59049 \sqrt{3}\kappa r_g}
\\\nonumber&&
+\frac{\imo \K \left(6599-940 \K^2\right)}{23328 \sqrt{3}\kappa^2 r_g}
-\frac{17 \imo \K\left(1580 \K^2-10087\right)\sigma^3}{209952 \sqrt{3}\kappa^2 r_g}
\\\nonumber&&
+\frac{\imo\K\left(38900\K^2+257903\right)\sigma^6}{177147 \sqrt{3}\kappa^2 r_g}
+\frac{854160\K^4-8009976\K^2-20776811}{60466176 \sqrt{3}\kappa^3 r_g}
\\\nonumber&&
+\frac{\left(58931760\K^4-330771624\K^2+114880295\right)\sigma^3}{1632586752 \sqrt{3}\kappa^3 r_g}
\\\nonumber&&
+\frac{\left(-3934320\K^4+50279448\K^2+23014913\right)\sigma^6}{28697814 \sqrt{3}\kappa^3 r_g}
\\\nonumber&&
-\frac{\imo\K\left(595922160\K^4+16670562456\K^2-30752137285\right)\sigma^3}{117546246144 \sqrt{3}\kappa^4 r_g}
\\\nonumber&&
-\frac{\imo\K\left(7185555888\K^4-33627171720\K^2+31238756407\right)\sigma^6}{99179645184 \sqrt{3}\kappa^4 r_g}
\\\nonumber&&
+\frac{\imo\K\left(11273136\K^4-258040200\K^2+1541370007\right)}{4353564672 \sqrt{3}\kappa^4 r_g}
+\Order{\kappa^{-5},\sigma^9}.
\end{eqnarray}

\begin{table}
\begin{flushright}
\begin{tabular}{|r|c|c|}
\hline
 $\sigma/\mu$& analytic formula & accurate value \\
\hline
  $-0.4$&$0.787729 - 0.171639 \imo$&$0.785798 - 0.172238 \imo$ \\
  $-0.2$&$0.752226 - 0.177035 \imo$&$0.751481 - 0.177594 \imo$ \\
  $   0$&$0.748072 - 0.177342 \imo$&$0.747343 - 0.177925 \imo$ \\
  $ 0.2$&$0.744069 - 0.177586 \imo$&$0.743356 - 0.178193 \imo$ \\
  $ 0.4$&$0.719281 - 0.178133 \imo$&$0.718889 - 0.178881 \imo$ \\
  $ 0.6$&$0.668247 - 0.175351 \imo$&$0.671440 - 0.176435 \imo$ \\
\hline
\end{tabular}
\end{flushright}
\caption{Dominant ($\K=1/2$) quasinormal modes of the black hole in the Effective Field Theory in the units of double mass ($\mu\cdot\omega$) for $\ell=2$ ($\kappa=5/2$), obtained using the analytic formula, and the accurate values for comparison.}\label{tabl:QNMSEFT}
\end{table}

In Table~\ref{tabl:QNMSEFT} we compare the dominant quasinormal modes, calculated using the analytic formula~(\ref{Hayward-beyondeikonal}), and the accurate values calculated in \cite{Mukohyama:2023xyf}. We see that for small values of the parameter $\sigma$ the analytic formula gives a good estimation of the frequencies even for small values of the multipole number $\ell$. Thus, we have a simple and robust method for the qualitative and quantitative analysis of the dependence of the quasinormal modes on the small parameter $\sigma$. In the shared Mathematica\textregistered{} notebook we present an automatic code,\footnote{The Mathematica\textregistered{} notebook is available from \url{https://arxiv.org/src/2309.02560v1/anc}.} which can be used, in a similar manner, in order to obtain approximate analytic formulas for the general spherically symmetric black-hole metrics.

\section{Transmission coefficients}\label{sec:greybody}

For the scattering problem we will consider the wave equation (\ref{wavelike}) under the boundary conditions that allow for incoming waves from infinity. Due to the symmetry of scattering properties, this is equivalent to the scattering of a wave originating from the horizon. Thus, the scattering boundary conditions are:
\begin{eqnarray}\label{reflBC}
\Psi = e^{-i\omega r_*} + R e^{i\omega r_*}, &\quad& r_*\to+\infty, \\
\label{transBC}
\Psi = T e^{-i\omega r_*}, &\quad& r_*\to-\infty,
\end{eqnarray}
where $R$ and $T$ are the reflection and transmission coefficients.

When solving the scattering problem with the help of the WKB method, the matching conditions allow us to express the reflection and transmission coefficients as follows \cite{Iyer:1986np},
\begin{eqnarray}\label{reflection}
|R|^2&=&\frac{1}{1+e^{-2\pi\imo \K}},\qquad 0<|R|^2<1.\\
\label{transmission}
|T|^2&=&\frac{1}{1+e^{2\pi\imo \K}}=1-|R|^2.
\end{eqnarray}
where $\K$ is a function of the frequency $\omega$, defined through the relation given by the WKB formula~(\ref{WKBformula-spherical}).

In order to obtain the reflection/transmission coefficients, equation~(\ref{WKBformula-spherical}) is usually solved numerically for a given $\omega$. Here we develop an analytic approach for calculation of $\K$ as a function of $\omega$ based on the series expansion in terms of $\kappa^{-1}$.

\subsection{Approximate analytic formula for the transmission coefficients}

We notice that the first-order WKB formula (\ref{WKBformula-spherical}) provides an analytic expression for $\K$,
\begin{equation}\label{eikonalK}
-\imo\K_0=\frac{\omega^2-V_0}{\sqrt{-2V_2}},
\end{equation}
which is the eikonal formula, providing the best accuracy, when the turning points are close,
\begin{equation}\label{deltaw}
\delta\equiv\omega^2-\Omega\kappa^2\approx0.
\end{equation}

At the same time it is clear that the nontrivial values of the reflection and transmission coefficients correspond to those values of $\omega$ for which the turning points are close and $\K\approx\K_0\approx0$.

We are now in position to obtain the expansion for $\K$ near the eikonal value $\K_0$. In order to obtain the corrections to the eikonal formula (\ref{eikonalK}), we must assume that $\K=\K_0+\Order{\kappa^{-1}}$.
Therefore, we imply that
\[\frac{\omega^2}{\kappa^2}-\Omega^2\equiv\frac{\delta}{\kappa^2}\approx-\imo\K_0\frac{\sqrt{-2V_2}}{\kappa^2}=\Order{\kappa^{-1}}.\]

Then, using the formula (\ref{WKBformula-spherical}) we calculate $\delta=\omega^2-\Omega\kappa^2$ and, from the obtained expression, find the expansion for $\K$ in terms of $\kappa$, considering $\delta=\Order{\kappa}$. In particular, for the Schwarzschild-de Sitter case, from (\ref{SdSspectrum}) after some algebra we find:
\begin{eqnarray}\label{KK}
\hspace{-30pt}-\imo\K&=&\frac{27M^2}{2 \s^2}\frac{\delta}{\kappa}+\frac{1-2s+2s^2}{4 \kappa}-\frac{137 - 216 s + 72 s^2}{432 \kappa}\s^2-\frac{81M^4\delta^2}{16\kappa^3}\frac{18 + 5 \s^2}{\s^4}
\nonumber\\&&\label{beyondeikonalK}
-\frac{81M^2\delta}{8\kappa^3}\left(67 - 105 s + 33 s^2-\left(\frac{9625}{144} - 105 s + 35 s^2\right)\s^2\right)
\nonumber\\&&
+\frac{M^6\delta^3}{\kappa^5}\frac{157464 + 43740 \s^2 + 31185 \s^4}{128\s^6}+\Order{\frac{1}{\kappa^3}},
\end{eqnarray}
where
\[\delta=\omega^2-\frac{\s^2}{27M^2}\kappa^2, \qquad \s\equiv\sqrt{1-9\Lambda M^2}.\]
With the above equation (\ref{KK}) for $\K$ at hand, and using eq.~(\ref{transmission}), one can immediately find the analytic approximation for the grey-body factors.

In order to compare the accuracy of the obtained grey-body factors we will compute the energy-emission rate which is usually the final aim of calculations of the transmission coefficients.
It is usually assumed that the black hole is in a state of thermal equilibrium with its surroundings in the sense that the temperature of the black hole remains constant between the emission of two consecutive particles, which is the canonical ensemble extensively discussed in the literature (see, e.g., \cite{Kanti:2004nr}). Then, the Hawking formula for the energy emission rate \cite{Hawking:1975vcx} can be applied:
\begin{equation}\label{energy-emission-rate}
\frac{d E}{d t} = \sum_{\ell}^{} N_{\ell} \intop_0^{\infty} |T_\ell|^2 \frac{\omega} {\exp\left(\omega/T_{H}\right)\pm1}\cdot \frac{d \omega}{2 \pi},
\end{equation}
where ``-'' sign is applied to the bosons ($s=0,1,2$) and ``+'' for the Dirac fermions ($s=1/2$).
Here $T_H$ is the Hawking temperature, $T_\ell$ are the transmission coefficients (grey-body factors) for the particular value of $\ell$, and $N_\ell$ are the corresponding multiplicity factors which depend on the number of species of particles. For the photons and gravitons we take into account two polarizations and $2\ell+1$ possible azimuthal numbers, so that
\begin{equation}\label{multiplicity}
N_{\ell} = 2(2\ell+1)=4\kappa.
\end{equation}
For the fermion particles, following \cite{Page:1976df}, we take into account two kind of particles and the corresponding antiparticles, so that
\begin{equation}\label{multiplicity-Dirac}
N_{\ell} = 4(2\ell+1)=8\kappa.
\end{equation}

\begin{table}
\begin{flushright}
\begin{tabular}{|l|l|l|l|l|l|}
\hline
 $i$&$\ell=1/2$ &$\ell=3/2$&$s=1$, $\ell=1$&$s=1$, $\ell=2$&$s=2$, $\ell=2$\\
\hline
 $0$&$2.21728\cdot10^{-4}$&$1.70175\cdot10^{-5}$&$0.47370\cdot10^{-4}$&$2.69999\cdot10^{-6}$&$2.69999\cdot10^{-6}$\\
 $1$&$1.81963\cdot10^{-4}$&$0.82303\cdot10^{-5}$&$0.55185\cdot10^{-4}$&$1.15825\cdot10^{-6}$&$8.98805\cdot10^{-6}$\\
 $2$&$1.70283\cdot10^{-4}$&$0.68464\cdot10^{-5}$&$0.41941\cdot10^{-4}$&$0.80659\cdot10^{-6}$&$4.68510\cdot10^{-6}$\\
 $3$&$1.58995\cdot10^{-4}$&$0.64161\cdot10^{-5}$&$0.38912\cdot10^{-4}$&$0.73074\cdot10^{-6}$&$4.68220\cdot10^{-6}$\\
 $4$&$1.60769\cdot10^{-4}$&$0.62439\cdot10^{-5}$&$0.36762\cdot10^{-4}$&$0.70049\cdot10^{-6}$&$4.19295\cdot10^{-6}$\\
 $5$&$1.51930\cdot10^{-4}$&$0.61462\cdot10^{-5}$&$0.35670\cdot10^{-4}$&$0.68710\cdot10^{-6}$&$4.09042\cdot10^{-6}$\\
 $6$&$1.57652\cdot10^{-4}$&$0.61057\cdot10^{-5}$&$0.35031\cdot10^{-4}$&$0.68002\cdot10^{-6}$&$4.00457\cdot10^{-6}$\\
\hline
Page&$1.575\cdot10^{-4}$&$0.60\cdot10^{-5}$&$0.330\cdot10^{-4}$&$0.7\cdot10^{-6}$&$3.8~\cdot10^{-6}$\\
\hline
\end{tabular}
\end{flushright}
\caption{The energy emission rate ($M^2\frac{dE}{dt}$) of the Schwarzschild black hole ($\s=1$) for the Standard Model fields and gravitons, calculated using the approximate formula for the transmission coefficient of the order $i$ in terms of $\kappa^{-1}$, and the accurate (Page) values for comparison.}\label{tabl:EER}
\end{table}

In Table~\ref{tabl:EER} we compare our approximate results for the Schwarzschild black hole ($\s=1$) with the values calculated by Don Page \cite{Page:1976df} and see that the energy-emission rate, calculated with the six-order approximation formula, provides quite good estimations even for small values of $\ell$ ($\kappa$).

In a similar fashion, the energy emission rates can be easily generalized to the case of non-zero cosmological constant. However, as Hawking radiation is primarily interesting for relatively small black holes, one can usually neglect cosmological factors and our focus here is on asymptotically flat black holes. We also omit consideration of grey-body factors for a test scalar field, because it does not contribute into the range of validity of the Standard Model considered in seminal works on Hawking radiation \cite{Page:1976df,Page:1976ki}.

\subsection{Further improvement of the calculation of the transmission coefficients}

From the table~\ref{tabl:EER} one can notice that the energy-emission rate has a systematic error, which makes our estimations larger than the accurate values. The reason for that is that the approximate formula (\ref{beyondeikonalK}) becomes inaccurate for small values of $\omega$, when the turning points go apart and the WKB approximation deteriorates. That is why the formula usually gives a nonzero transmission coefficient even in the infrared limit ($\omega\to0$). The obtained ``contribution'' to the energy-emission rate is not negligibly small, leading to the systematic overestimation of the total rate. The same is true for the ultraviolet radiation ($\omega\gtrsim M^{-1}$), even though it is exponentially suppressed and, therefore, not so important for the calculation of the integral (\ref{energy-emission-rate}).

Although suggesting a robust amendment of this problem is beyond the scope of the present work, we will discuss some approaches for the practical elimination of the systematic errors of the infrared and ultraviolet radiation.

\begin{figure}
\resizebox{\linewidth}{!}{\includegraphics{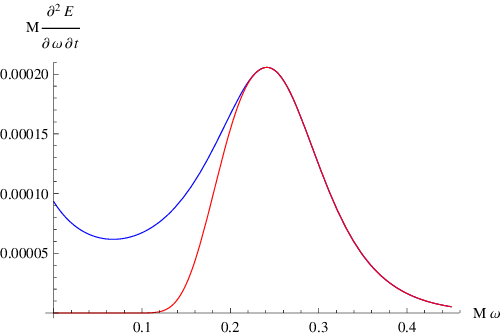}}
\caption{Energy emission spectrum rate of the Schwarzschild black hole ($\s=1$) owing to the photons ($s=1$) at $\ell=1$, calculated using the second-order formula (\ref{beyondeikonalK}) without the extrapolation (blue) and using of the extrapolation (red).}\label{fig:EERimproved}
\end{figure}

First we recall that the formula (\ref{beyondeikonalK}) provides a good approximation when the higher-order terms are small corrections to the eikonal formula (\ref{eikonalK})
\begin{equation}
-\imo\K=-\imo\K_0+\Order{\frac{1}{\kappa}}=\frac{\omega^2-V_0}{\sqrt{-2V_2}}+\Order{\frac{1}{\kappa}},
\end{equation}
which is a monotonously growing function of $\omega>0$, and its second derivative is a positive constant.

This allows us to formulate a simple test for the validity of the formula (\ref{beyondeikonalK}): We assume that the approximation for the transmission coefficient is valid in the interval, containing $\omega=\Omega$, between the two points where either first or second derivative of $\K$ with respect to $\omega$ vanish. Outside this interval we will approximate the transmission coefficient $T(\omega)$ by its infrared or ultraviolet asymptotics. We consider the simplest form of such asymptotic behaviours, inspired by Eq.~\ref{transmission}:
\begin{equation}\label{transmission-asymtotic}
\begin{array}{rclcl}
|T|^2&\approx& Ae^{-a/\omega^2}, &\quad& \omega\ll\Omega,\\
|T|^2&\approx& 1-Be^{-b\cdot\omega^2}, &\quad& \omega\gg\Omega,\\
\end{array}
\end{equation}
where $A$, $a$, $B$, $b$ are positive constants, which are obtained by matching the value of $|T|^2$ and its derivative at the ends of the above interval. If any of the derivatives is zero, we set the factor $A=1$ or $B=1$ and match only the value of $|T|^2$ with the asymptotic formula (\ref{transmission-asymtotic}) at the corresponding endpoint, in order to have negative argument in the exponent.

The described above amendment significantly improves the accuracy of the energy-emission rate and its convergence in terms of the approximation order. In Fig.~\ref{fig:EERimproved} we show the result of the extrapolation of the transmission coefficient in the infrared region. We see that the low-energy photons, which cannot tunnel through the effective potential indeed do not contribute to the emission spectrum when we use the extrapolation (\ref{transmission-asymtotic}).

\begin{table}
\begin{flushright}
\begin{tabular}{|l|l|l|l|l|l|}
\hline
 $i$&$\ell=1/2$ &$\ell=3/2$&$s=1$, $\ell=1$&$s=1$, $\ell=2$&$s=2$, $\ell=2$\\
\hline
 $0$&$2.21728\cdot10^{-4}$&$1.70175\cdot10^{-5}$&$0.473704\cdot10^{-4}$&$2.69999\cdot10^{-6}$&$2.69999\cdot10^{-6}$\\
 $1$&$1.36150\cdot10^{-4}$&$0.53239\cdot10^{-5}$&$0.319793\cdot10^{-4}$&$0.63253\cdot10^{-6}$&$1.55666\cdot10^{-5}$\\
 $2$&$1.42727\cdot10^{-4}$&$0.55711\cdot10^{-5}$&$0.289071\cdot10^{-4}$&$0.61411\cdot10^{-6}$&$3.20007\cdot10^{-6}$\\
 $3$&$1.46414\cdot10^{-4}$&$0.58156\cdot10^{-5}$&$0.309669\cdot10^{-4}$&$0.64969\cdot10^{-6}$&$3.52029\cdot10^{-6}$\\
 $4$&$1.49914\cdot10^{-4}$&$0.59236\cdot10^{-5}$&$0.318378\cdot10^{-4}$&$0.66124\cdot10^{-6}$&$3.59602\cdot10^{-6}$\\
 $5$&$1.50385\cdot10^{-4}$&$0.59678\cdot10^{-5}$&$0.325098\cdot10^{-4}$&$0.66580\cdot10^{-6}$&$3.74835\cdot10^{-6}$\\
 $6$&$1.52275\cdot10^{-4}$&$0.59890\cdot10^{-5}$&$0.327605\cdot10^{-4}$&$0.66736\cdot10^{-6}$&$3.79877\cdot10^{-6}$\\
\hline
Page&$1.575\cdot10^{-4}$&$0.60\cdot10^{-5}$&$0.330\cdot10^{-4}$&$0.7\cdot10^{-6}$&$3.8~\cdot10^{-6}$\\
\hline
\end{tabular}
\end{flushright}
\caption{The energy emission rate ($M^2\frac{dE}{dt}$) of the Schwarzschild black hole ($\s=1$) for the Standard Model fields and gravitons, calculated using the approximate formula for the transmission coefficient of the order $i$ in terms of $\kappa^{-1}$ with extrapolation improvement, and the accurate (Page) values for comparison.}\label{tabl:EERimproved}
\end{table}

In Table~\ref{tabl:EERimproved} we show the energy-emission rates calculated with the help of the extrapolation improvement. We see that the values generally converge faster and provide better approximation, comparing to the results obtained with the analytic formula of the same order for all values of $\omega$ without the extrapolation.

A tricky point here is again related with the asymptotic-only convergence of the WKB series. Occasionally, at some order of the WKB or $1/\ell$ expansions the results can be very close to the precise values. These cases must be distinguished from the plateau of relative convergence, when the results given, at the nearby, say, 4th, 5th and 6th, orders do not differ much. That last case is where the WKB formula makes its best. Thus, although the values for $i=6$ of the second and fifth columns of Table 6 are closer to the accurate values, this is, obviously, a numerical coincidence, which can be easily seen by comparing $i=5$ and $i=4$ values. In both cases the second decimal place varies in Table~\ref{tabl:EER}, suggesting that it cannot be accurately estimated without the extrapolation improvement. The corresponding values of Table~\ref{tabl:EERimproved} are much more numerically stable in that sense that they less sharply depend on a change of the expansion order. Since we suggest no universal algorithm for removing the systematic error of the WKB approximation, we believe that there is no need for a rigorous error analysis of our estimations.

One can see that the energy-emission rate for fermions is higher than that for bosons. There are two main physical reasons for that. First, the lower multipole number implies the lower tunneling barrier, which increases the probability of lower-energy particles to reach an observer. Since the multipole number starts from the spin parameter, the emission of fermions (s=1/2) are less suppressed by the grey-body factors. The second reason is the number of species and degrees of freedom for bosons and fermions. For the gravitational and electromagnetic field there are only two polarizations in the four-dimensional spacetime. If one studies the Hawking radiation in higher-dimensional space-time, where the number of gravitational degrees of freedom is larger, considerable or even dominant contribution may come from the emission of gravitons as well \cite{Kanti:2009sn}.

Finally we would like to stress that the improvement considered in this section is based on an ad hoc extrapolation of the transmission coefficients (grey-body factors) to the infrared and ultraviolet regimes. Since the analytic formula based on the WKB method provides a good approximation for the dominant range of frequencies of the energy-emission spectra, we have a significant improvement of the accuracy, once we take into account the correct asymptotic behaviour of the transmission coefficients. However, a more careful treatment of the infrared and ultraviolet behaviour of the transmission coefficients must be used if one needs to study the grey-body factors in these regimes.

An alternative approach could use the Padé approximants in terms of $\delta$ and/or $\kappa^{-1}$ rather than polynomial series. Since the Padé approximants are useful for the higher-order WKB formula for quasinormal modes (see \cite{Konoplya:2019hlu} for a review) we believe that the development of the grey-body factor analytic formula in terms of the appropriate Padé approximants might improve its accuracy and the infrared/ultraviolet behaviour.

\section{Conclusions}

The analytical treatment of spectral problems concerning black holes undoubtedly holds an advantage over purely numerical computations. Quasinormal modes and grey-body factors associated with Schwarzschild black holes and their generalizations can only be ascertained numerically due to the intricate nature of the master wave equation. The eikonal formula, offering reasonable accuracy only for moderate and high values of the multipole number $\ell$, remains a rare instance of such analytical treatment. In this context, we have introduced a comprehensive approach to analytically treat quasinormal modes and grey-body factors beyond the eikonal limit. This approach is presented as an expansion in powers of $1/\ell$, necessitating the simultaneous utilization of higher-order WKB expansions.

This general procedure has been exemplified through two instances: Schwarzschild-de Sitter black holes and a hairy black hole within the Effective Field Theory. In both cases, concise analytical formulas were derived for quasinormal modes and grey-body factors. For the latter, the eikonal formula has been obtained for the first time, to the best of our knowledge. Despite the asymptotic convergence nature of the WKB series, the resulting analytical expressions yield a high degree of accuracy across all values of~$\ell$.

The prospects of our work extend to axially symmetric black holes, at least in scenarios where the perturbation equations permit the separation of angular variables \cite{Konoplya:2018arm, Papadopoulos:2018nvd}. A notable enhancement in accuracy could be achieved by including Padé approximants \cite{Matyjasek:2017psv} into the $1/\ell$ expansion procedure.

\section*{Acknowledgments}
A.~Z. was supported by Conselho Nacional de Desenvolvimento Científico e Tecnológico (CNPq).

\bigskip

\bibliographystyle{unsrt}
\bibliography{Bibliography}

\end{document}